\documentstyle[12pt]{article}

\newcommand{\pat}{\partial}
\newcommand{\be}{\begin{equation}}
\newcommand{\ee}{\end{equation}}
\newcommand{\bea}{\begin{eqnarray}}
\newcommand{\eea}{\end{eqnarray}}
\newcommand{\abf}{{\bf a}}

\newcommand{\zcal}{z_{12}}
\newcommand{\Acal}{{\cal A}}
\newcommand{\Fcal}{{\cal F}}
\newcommand{\half}{\frac{1}{2}}
\newcommand{\nbyn}{N \times N}
\newcommand{\repres}{\leftrightarrow}
\newcommand{\tr}{{\rm Tr}}
\newcommand{\ninfty}{N \rightarrow \infty}
\newcommand{\unitk}{{\bf 1}_k}
\newcommand{\unittwo}{{\bf 1}_2}
\newcommand{\holo}{{\cal U}}
\newcommand{\bra}{\langle}
\newcommand{\ket}{\rangle}
\newcommand{\muhat}{\hat{\mu}}
\newcommand{\nuhat}{\hat{\nu}}
\newcommand{\rhat}{\hat{r}}
\newcommand{\shat}{\hat{s}}

\begin{document}

\baselineskip 19pt

\begin{titlepage}
\begin{flushright}
CALT-68-2123 \\
June 1997
\end{flushright}

\vskip 1.0truecm

\begin{center}
{\Large {\bf Notes on Branes in Matrix Theory}}
\end{center}

\vskip 0.6cm

\begin{center}
{\bf Esko Keski-Vakkuri}$^1$ and {\bf Per Kraus}$^2$
\vskip 0.3cm
{\it California Institute of Technology \\
     Pasadena CA 91125, USA  }
\end{center}

\vskip 2.0cm

\begin{center}
{\small {\bf Abstract: }}
\end{center}       
\noindent           
{\small We study the effective actions of various 
brane configurations in Matrix theory. 
Starting from the 0 + 1 dimensional quantum mechanics, we replace 
coordinate matrices by covariant derivatives in the large N limit, thereby 
obtaining effective field theories on the brane world volumes. Even 
for noncompact branes, these effective theories are of Yang-Mills type,
with constant background magnetic fields.  In the case of a D2-brane, we
show explicitly how the effective action equals the large magnetic field
limit of the Born-Infeld action, and thus derive from Matrix theory the action
used by Polchinski and Pouliot to compute M-momentum transfer between 
membranes.  
We also consider the effect of 
compactifying transverse directions. Finally, we analyze a scattering 
process involving a recently proposed background representing a classically
stable D6+D0 brane configuration. 
We compute the potential between this configuration and a D0-brane, 
and show that the result agrees with supergravity.
}
\rm
\vskip 2.4cm

\small
\begin{flushleft}
$^1$ Work supported in part by a DOE grant DE-FG03-92-ER40701. \\ 
E-mail: {\em esko@theory.caltech.edu} \\
$^2$ Work supported in part by a DOE grant DE-FG03-92-ER40701 and
by a DuBridge Fellowship. E-mail: {\em perkraus@theory.caltech.edu}
\end{flushleft}
\normalsize
\end{titlepage}

\newpage
\baselineskip 22pt

\section{Introduction}

Matrix theory \cite{BFSS} purports to be a complete, non-perturbative 
formulation of M theory. Since a great deal is known about the 
physics of M theory in various corners of moduli space, much 
effort has recently been directed towards verifying that such 
physics can be recovered from Matrix theory. In this paper we will 
be primarily interested in studying the description of D-branes that 
emerges when Matrix theory is compactified on a circle, yielding the 
type IIA theory. Some salient features of D-branes in IIA theory which 
one would like to obtain from Matrix theory include the following: there 
exist Dp-branes for p = 0,2,4,6,8; the low energy dynamics of k parallel 
Dp-branes is described by a supersymmetric p+1 dimensional U(k) gauge 
theory; the (not necessarily low energy) dynamics of a single D-brane is 
described by a Born-Infeld type action. In addition, one knows the 
supergravity fields produced by D-branes and so also the corresponding 
small angle scattering amplitudes of several such objects.

Banks, Seiberg, and Shenker \cite{BSS}, and S.-J. Rey \cite{SJR}, 
took some first steps towards 
deriving the effective actions of D-branes from Matrix theory. They 
identified matrix assignments corresponding to D-branes by identifying 
the appropriate central charges in the 11 dimensional light cone gauge 
supersymmetry algebra. Then by analyzing the fluctuations around such 
backgrounds and replacing large N commutators by Poisson brackets, they 
were able to recover the general structure of the known effective actions 
to quadratic order in fields. Here, our goal is similar, but we wish to go 
beyond quadratic order and to keep careful track of numerical factors. To 
do so successfully, we found it necessary to replace matrix commutators not 
by Poisson brackets but rather by the commutators of covariant derivatives, 
much as one does when considering toroidal compactification. By so doing, 
we are able to obtain gauge invariant actions which precisely reproduce
the  fluctuations of the Born-Infeld action in the presence of a 
large background magnetic field.
	
There have been a number of studies comparing D-brane scattering 
amplitudes obtained from Matrix theory and from supergravity. On the Matrix 
theory side, the procedure involved computing the effective action of a 
finite N matrix configuration, and then extrapolating to large N. Strictly 
speaking, this procedure is inconsistent as the backgrounds require that 
certain finite N matrices have a commutator proportional to the identity, 
even though such matrices cannot exist. In our approach, we avoid this 
problem by taking $\ninfty$ from the outset. As a specific application, 
we consider the scattering of a D0-brane from a configuration
composed purely of D6 and D0 branes \cite{WT}, which has not 
been studied before. We find 
complete agreement at the one loop level between Matrix theory and 
supergravity.

The remainder of this paper is organized as follows. In section 2 
we rewrite the Matrix theory action in a manner suitable for the study of 
D-branes. Sections 2.1 and 2.2 focus on the 
effective action for D2-branes, and 
section 2.3 on the effective action for D4-branes. In section 3 we compute the 
potential between a D0-brane and a D6+D0 configuration, and in section 4 we 
discuss some of our results.

\bigskip

\section{Fluctuations of branes in Matrix theory}

The action governing Matrix theory is gotten by 
dimensionally reducing the action of 10 dimensional 
Super Yang-Mills theory to 0+1 dimensions.
In string units ($2\pi \alpha' =1$) the 
bosonic part of the Lagrangian is\footnote{following \cite{BFSS}, Appendix B}
\be
  {\cal L}_B = \tr \ L_B \ ; \ L_B = \frac{T_0}{2}
 \{ (D_0X_I)^2 + \frac{1}{2} [X_I,X_J]^2 \} 
\ee
where $T_0$ is the D0-brane ``tension''
$$
T_0 = \frac{1}{\sqrt{\alpha'} g} = \frac{\sqrt{2\pi}}{g} \ ,  
$$
the covariant derivative $D_0$ is
$$
D_0 = \pat_t - i[A_0,\ \cdot \  ]
$$
and $I=1,2,\ldots ,8,9$. A p-brane background is described by
\bea
   X_r &=& U_r \ ; \  r=1,2,\ldots ,p \\ 
   X_i &=& 0   \ ; \ i=p+1,\ldots ,9  
\eea
where $U_r, r=1,\ldots ,p$ are certain matrices to 
be specified later.

Fluctuations around this background are denoted by $A_r,\phi_i$,
\be
  X_r = U_r + A_r
\label{axar}
\ee
\be 
  X_i = \phi_i \ .
\ee
Following Banks, Seiberg and Shenker \cite{BSS}, we
substitute the above expansions into the Lagrangian $L_B$, and
then regroup and rename the terms so that we arrive at the following
form:
\be
L_B = \frac{T_0}{2} \{ -\frac{1}{2} (F_{\mu \nu})^2 + (D_{\mu}\phi_i)^2
      +\frac{1}{2} [\phi_i,\phi_j ]^2 + i[U_r,U_s]F_{rs} 
          + \half [U_r,U_s]^2 \} \ ,
\label{basicL}
\ee
where $\mu,\nu = 0,1,\ldots ,p; r,s=1,\ldots ,p$ and
\begin{eqnarray*}
 F_{0r} &=& -F_{r0} = \pat_0 A_r + i[U_r,A_0] - i[A_0,A_r] \\
 F_{rs} &=& -i[U_r,A_s] + i[U_s,A_r] -i[A_r,A_s] \\
 D_0\phi_i &=& \pat_0\phi_i -i[A_0,\phi_i] \\
 D_r\phi_i &=& -i[U_r,\phi_i] -i[A_r,\phi_i] \ . 
\end{eqnarray*}
(In the above, $(D_{\mu}\phi_i)^2 = (D_0\phi_i)^2 - (D_r\phi_i)^2$ etc.)

So far, we have not done anything other than reorder the terms and
give them new names. Now we try to interpret the result
in the context of some particular p-brane backgrounds. 

\bigskip

\subsection{Infinite membrane backgrounds (p=2)} 

\bigskip

The 11 dimensional SUSY algebra admits central charges corresponding to the 
presence of membranes. Banks, Seiberg and Shenker \cite{BSS} computed the
central charges starting from the Matrix theory action 
and found\footnote{with the normalization corresponding 
to conventions in this paper}
\be
    Z_{rs} = -\frac{i}{2\pi} \tr [X_r,X_s] \ .
\ee
It follows immediately that $Z_{rs}$ vanishes for finite $N$. On the other
hand, it has recently been proposed by Susskind \cite{Nfin} that the
finite $N$ version of Matrix theory is to be interpreted as describing a 
finite longitudinal momentum sector of M theory quantized in light cone
gauge. The vanishing of $Z_{rs}$ at finite $N$ can be understood in this
light, for a membrane carrying finite momentum must necessarily be compact
and so carry no net central charge. 

Ref. \cite{BFSS} discussed a formal 
method for constructing membrane backgrounds in the $\ninfty$ limit. The
construction uses
canonical variables $Q,P$ obeying the commutation relation
\be
     [Q,P] = \frac{2\pi i}{N} \ .
\ee
Since this relation cannot be satisfied at finite $N$, we will instead
take the $\ninfty$ limit in terms of the rescaled variables \cite{AB}
\be
    U_1 = Q\sqrt{N\zcal } \ ; \ U_2 = P\sqrt{N\zcal }  \ .
\ee
Taking the spectrum of $P,Q$ to go from 0 to $2\pi$, the area of the membrane
is $A = (2\pi )^2 N \zcal$, becoming infinite in the $\ninfty$ limit.
The commutation relation
\be
  [U_1,U_2] = 2 \pi i \zcal
\label{infbrane}
\ee
can be represented by differential operators acting on a space of functions.
More formally, the $N$ dimensional vector space $V_N$ on which the matrices
$X$ act will be replaced by an infinite dimensional space 
$V_{\infty}$ of functions on the membrane worldvolume. The membrane background
$U_1,U_2$ is represented by differential operators acting on $V_{\infty}$,
and fluctuations will be represented by elements of $V_{\infty}$,
{\em i.e.}, functions on the membrane worldvolume.

A simple representation of $U_{1,2}$ found in the literature 
consists of taking
\be
   U_1 = 2\pi i \zcal \pat_y \ \ ; U_2 = y
\ee
in the $N\rightarrow \infty$ limit. As Aharony and Berkooz \cite{AB} have 
pointed out, a cosmetic flaw of this 
representation is that a membrane  
carries two spatial coordinates whereas only one is
manifest in the above equation. 

There exists, however, a more ``natural'' representation than
the one given above. Membranes carry a charge, which
in Matrix theory 
arises from a U(1) subgroup of the U(N) symmetry, or in the continuum
limit, the U(1) subgroup of U($\infty $) which is isomorphic with the 
infinitesimal area preserving diffeomorphisms of the membrane.
Specifically, the membrane charge density $2\pi \zcal$ is 
associated with a U(1) magnetic field living on the membrane world volume.
In ten dimensions the boosted membrane corresponds to a D2-brane with a 
large background magnetic field $f_{12}$ on its worldvolume, related to
the density $\sigma_0$ of D0-branes bound to the D2-brane by \cite{Town}:
\be
       \sigma_0 = \frac{1}{2\pi} f_{12} \ .
\ee
Since the D0-brane density is 
\be
       \sigma_0 = \frac{N}{A} =   \frac{1}{(2\pi )^2 \zcal} \ ,
\label{d0density}
\ee
we find that the magnetic field is given by
\be
           f_{12} = \frac{1}{2\pi \zcal} \ .
\label{magnfield}
\ee
Let $\abf \equiv (a_1,a_2)$ be the background U(1) vector potential 
corresponding to the magnetic field: $f_{12} = \nabla \times \abf$. We choose
\be
    a_1 = -\half f_{12} x_2 \ \ , \ \ a_2 = \half f_{12} x_1  \ \ .
\label{backgr}
\ee
Then, a representation of $U_{1,2}$, in which both of the membrane 
coordinates and the background magnetic field are manifest, is given by
\be \begin{array}{l}
  U_1 \leftrightarrow 2\pi \zcal \ (i\pat_{x_1} + a_1(x_1,x_2)) \\ 
  U_2 \leftrightarrow 2\pi \zcal \ (i\pat_{x_2} + a_2(x_1,x_2)) \end{array} \ .
\label{covder}
\ee
This representation satisfies the commutation relation (\ref{infbrane}).
To avoid confusion, let us point out that although expressions 
similar to (\ref{covder}) appear in
the context of toroidal compactification \cite{GRT}, 
here the motivation and the interpretation 
are quite different. 
We have not T-dualized anything,
the two coordinates $x_1,x_2$ parametrize the infinite membrane, 
$(x_1,x_2) \in R^2$.   

All other $\nbyn $ matrices are represented by functions on the membrane 
worldvolume:
\bea
 \phi_i &\leftrightarrow & \phi_i (x_1,x_2) \\
    A_0 &\leftrightarrow & A_0(x_1,x_2) \nonumber \ .
\label{others}
\eea
(We have suppressed the dependence on the worldvolume time coordinate.)
For the components $A_r,\ r=1,2$ we will choose
\be
  A_r \leftrightarrow 2\pi \zcal \  A_r(x_1,x_2)  \ .
\label{aar}
\ee
With this normalization, (\ref{axar}) becomes
\be
 X_r \leftrightarrow 2\pi \zcal \ [i\pat_{x_r} + a_r(x_1,x_2) 
                                               + A_r(x_1,x_2)] \ ,
\ee
and we see that $A_{\mu}(x_1,x_2)$ then has the proper interpretation
as the fluctuation around the background (\ref{backgr}).

In the $\ninfty$ limit, the trace operation is replaced by  
integration over the spatial membrane worldvolume coordinates $x_1,x_2$:
\be
     \tr \leftrightarrow \sigma_0 \int \ dx_1dx_2 \ \ ,
\label{trace1}
\ee
where the normalization factor
$\sigma_0$ is included in 
order to preserve the interpretation of the rank
of the matrices as the number of D0-branes $N=Tr {\bf 1}_N$; we
represent the unit matrix ${\bf 1}_N$ by 1.

Finally, the matrix commutators 
which involve $U_r$ become commutators of operators
and functions, {\em e.g.}
\be
     [U_r,A_0] \leftrightarrow (2\pi \zcal)  
                              [i\pat_r + a_r(x_1,x_2) , A_0(x_1,x_2)] \ ,
\ee
and the matrix commutators which do not involve $U_r$ become commutators
of functions, {\em e.g.}
\be
     [\phi_i,\phi_j] \leftrightarrow  [\phi_i(x_1,x_2),\phi_j(x_1,x_2)] \ .
\label{commutator}
\ee
In the case of a single membrane, these will be zero. 

Now we can see what happens to $L_B$. It becomes a 2+1 dimensional
Lagrangian density with the terms
$$
\begin{array}{lcccr}
 F_{0r} & = & \pat_0 A_r +i[U_r, A_0] - i[A_0,A_r]
    & \leftrightarrow & (2\pi \zcal )\ (\pat_0 A_r(x) - \pat_r A_0(x)) \\ 
F_{12} & = &  -i[U_1,A_2] +i[U_2,A_1] - i[A_1,A_2]
  & \leftrightarrow & (2\pi \zcal )^2 \  (\pat_1 A_2(x) - \pat_2 A_1(x)) \\
D_0\phi_i & = & \pat_0 \phi_i - i[A_0,\phi_i] 
 & \leftrightarrow & \pat_0 \phi_i(x) \\
D_r\phi_i & = &  -i[U_r,\phi_i] - i[A_r,\phi_i] 
 & \leftrightarrow & (2\pi \zcal ) \ (\pat_r \phi_i (x)) 
\end{array} 
$$
We find
\be
  {\cal L}_B = \tr L_B \leftrightarrow
    \sigma_0 \int d^2x \ \frac{T_0}{2} \{ -\half 
     (F_{\mu \nu} F^{\muhat \nuhat}) + 
    \pat_{\mu} \phi_i \pat^{\muhat} \phi_i 
    -2 (2\pi \zcal )^3 F_{12} - (2\pi \zcal )^2 \} \ ,
\label{U(1)L}
\ee
where $F_{\mu \nu} = \pat_{\mu} A_{\nu} - \pat_{\nu} A_{\mu}$, and
the hatted upper indices are raised with the metric
\be
 \hat{\eta}_{\mu\nu} = {\rm diag}\ (1, -\frac{1}{(2\pi \zcal )^2}, 
                                     -\frac{1}{(2\pi \zcal )^2}) \ .
\label{ppmetric}
\ee 
As a check, using the last term we can work out the tension of the membrane
to be 
\be
    T_2  = \frac{1}{2\pi} T_0
\label{tension}
\ee
which is correct.
Notice that a key feature of our representation 
in the $\ninfty$ limit is the vanishing of commutators such
as (\ref{commutator}). This is in contrast to the approach in \cite{BSS}, 
where the authors represented the commutators of matrices
by Poisson brackets. Such a representation yields  nonvanishing
contributions such as 
$$
  \{A_r,A_s \} 
$$
to the $F_{rs}$ component of the field strength, which is cumbersome from 
the D2-brane point of view where one expects to obtain
a U(1) symmetry. This puzzle seems to be related to the distinction 
between infinite and finite membranes; for 
discussion, see \cite{AB}.
Note also that even if the commutators vanish for a single infinite
membrane, we will obtain nonvanishing commutators   
later when we discuss the case of multiple membrane backgrounds.

It is interesting to examine the relation between the effective action
(\ref{U(1)L}) which was derived from Matrix theory, and the analogous
Born-Infeld action for a D2-brane in IIA theory. The latter is
\be 
  L_{BI} = -T_2 \int \! d^2x \sqrt{\det [\eta_{\mu \nu} + F_{\mu \nu}
-\partial_\mu \phi_i \partial_\nu \phi_i]} \ ,
\label{BI}
\ee
We recall that boosting to the infinite momentum frame in the eleven 
dimensional description corresponds in ten dimensions to turning on a large
magnetic field on the worldvolume of the D2-brane.  We therefore make the
replacement $F_{12} \rightarrow f_{12}+F_{12}$, where 
$f_{12} = {1 \over {2\pi z_{12}}}$ as in (\ref{magnfield}), and 
expand $L_{BI}$ for large $f_{12}$. To quadratic order in fluctuations we find
\bea
 L_{BI} &=&  -T_2 \int \! d^2x \left\{ \frac{1} {2 \pi z_{12}} 
                                      +F_{12} \right\} \\
\mbox{} &-& {T_2 \over 2} \int \! d^2x  \left\{ (2\pi z_{12})
 + (2\pi z_{12})^2 F_{12} + {1 \over (2\pi z_{12})} {1\over 2} 
           F_{\mu\nu}F^{\hat{\mu}\hat{\nu}} 
     - {1 \over (2\pi z_{12})} \pat_{\mu} \phi_i \pat^{\muhat} \phi_i\right\}
       \nonumber \ .
\label{expandBI}
\eea
Again, upper indices are raised with the metric (\ref{ppmetric}).
To compare with Matrix theory the expression in the first set of  braces 
should be subtracted according to $H_{IMF}=H-p_{11}$.  Then, using 
$T_2=T_0/2\pi$ and (\ref{d0density}), comparison 
of $L_{BI}$ with (\ref{U(1)L}) shows complete agreement except
for a factor of $1/2$ multiplying the term linear in $F_{12}$.  This term is
a total derivative and does not appear to  be significant.

The action (27) represents the starting point for a calculation
of Polchinski and Pouliot \cite{PP} involving M-momentum transfer between 
membranes. (Related discussion can be found in \cite{DKM,BFSS2}.)
There it was assumed that the action could be obtained directly from Matrix
theory; here we have shown explicitly how this can be accomplished.

Consider now a background configuration of $k$ parallel
infinite membranes in the transverse directions $X_1,X_2$. At finite
$N$, this is represented by block diagonal $kN\times kN$ matrices, 
$X_1$ has $k$ copies of the ``$\nbyn $ matrix'' $Q$ in the diagonal, and $X_2$ 
has $k$ copies of $P$. Correspondingly, for $k$ infinite membranes
we take
\bea
     U_1 &\leftrightarrow & 2\pi \zcal (i\pat_1 + a_1) \otimes {\bf 1}_k \\
     U_2 &\leftrightarrow & 2\pi \zcal (i\pat_2 + a_2) \otimes {\bf 1}_k
\eea
where ${\bf 1}_k$ is the unit $k\times k$ matrix. Thus,
\be
   [U_1,U_2] = 2\pi i \zcal \unitk \ .
\label{kcharge}
\ee
This background breaks
the $U(\infty )$ symmetry down to an $U(k)$ symmetry. The fluctuations
$A_r,\phi_i$ are not block diagonal, which would correspond to fluctuations
within each brane, but also contain off diagonal components 
corresponding to strings connecting distinct membranes,
thereby accounting for the full $kN\times kN$ matrix structure.
The matrices can be represented as a sum of tensor
products of $\nbyn$ matrices and generators 
of the $U(k)$ Lie algebra \cite{BSS,SJR}. 
The $\nbyn$ parts, in the $\ninfty$ limit, are replaced by functions
on the membrane - thus (17), (18)
are now replaced by
\bea
A_0 &\leftrightarrow & A^0_0(x_1,x_2)\  \unitk + A^a_0(x_1,x_2)\  T^a  \\      
A_r &\leftrightarrow & 2\pi \zcal \ (A^0_r(x_1,x_2)\ \unitk 
                                + A^a_r(x_1,x_2) \ T^a ) \nonumber \\   
\phi_i &\leftrightarrow & \phi^0_i(x_1,x_2)\ \unitk + \phi^a_i(x_1,x_2)\ T^a 
    \nonumber \ .
\label{othersk}
\eea
where $T^a$ are the traceless generators of the $SU(k)$ Lie algebra.
Commutators thus become commutators of fields in the adjoint
representation of $U(k)$, {\em e.g.} (\ref{commutator}) becomes
$$
     [\phi_i(x_1,x_2) ,\phi_j(x_1,x_2) ]_k
$$
and the $SU(k)$ part will survive. 

The trace over $kN\times kN$ matrices is now represented as follows
\be
 \tr \repres \sigma_0 \int \ dx_1dx_2 \tr_k \ ,
\label{tracek}
\ee
where $\tr_k$ is the trace over the $k\times k$ structure. Note that
the total membrane charge of the multimembrane configuration is
\bea
   Z_{12} = -\frac{i}{2\pi} 
   \ Tr[X_1,X_2] &\repres & \sigma_0 \int dx_1dx_2 \zcal \tr_k \unitk
    \\ \nonumber
               &=& k \cdot {\rm (charge\ of\ a\ single\ membrane)} 
\eea
as it should be. Plugging everything into the Lagrangian (\ref{basicL}),
we obtain 
\be
    {\cal L}_B \repres
    \sigma_0 \int \ d^2 x  \ \frac{T_0}{2} \tr_k \ \{ -\half 
       (F_{\mu \nu} F^{\muhat \nuhat}) + D_{\mu} \phi_i D^{\muhat} \phi_i
     + \half ([\phi_i ,\phi_j ]_k)^2 
     - 2(2\pi \zcal )^3 F_{12}
     - (2\pi \zcal )^2 \ \unitk \ \}
\label{U(k)L}
\ee
where
$$
\begin{array}{rcl}
 F_{0r} & = & \pat_0 A_r(x) - \pat_r A_0(x) -i[A_0(x),A_r(x)]_k \\ 
 F_{12} & = & \pat_1 A_2(x) - \pat_2 A_1(x) -i[A_1(x),A_2(x)]_k \\
D_0\phi_i & = & \pat_0 \phi_i(x) -i[A_0(x),\phi_i(x)]_k \\
D_r\phi_i & =  & \pat_r \phi_i (x) -i[A_r(x),\phi_i(x)]_k
\end{array} 
$$
We can also check the gauge transformation properties directly.
The Matrix Lagrangian (\ref{basicL}) is invariant under the $U(kN)$ gauge
transformations 
\bea
   \delta A_r &=& -i[U_r,\lambda ] + i [\lambda, A_r ] \\ 
   \delta \phi_i &=& i[\lambda , \phi_i ] \nonumber \\
   \delta A_0 &=& \pat_0 \lambda + i[\lambda ,A_0] \nonumber
\eea
where the gauge transformation parameter $\lambda$ (and everything else) 
is a $kN\times kN$ matrix. In the $\ninfty$ limit these become
\bea
   \delta A_r(x) &\repres & \pat_r\lambda (x) + i [\lambda (x), A_r(x) ]_k \\ 
  \delta \phi_i(x) &\repres & i[\lambda (x) , \phi_i (x)]_k \nonumber \\
\delta A_0(x) &\repres & \pat_0 \lambda (x) +i[\lambda (x),A_0(x)]_k \nonumber
\eea
and give the correct $U(k)$ gauge symmetry of (\ref{U(k)L}).

\bigskip

\subsection{Wrapped membrane backgrounds (p=2)} 

\bigskip

We now compactify the Matrix theory on a two dimensional torus
$T^2$ in the $X_1,X_2$ directions.
This corresponds to compactifying M theory on $T^2 \times S^1$, since 
one spatial
direction was already compactified.
We will then consider backgrounds of transverse membranes which wrap around
the torus $T^2$. As a result of the compactification, there
will be additional degrees of freedom in the theory, due to strings which
wind around the compact directions. As explained by 
Taylor \cite{GRT}, they can be 
accomodated in the Matrix theory by replacing each
partonic D0-brane by a countably infinite number of copies of it on the
noncompact covering space of the torus. 
The vector space $V$,
on which the matrices act, then has a tensor product
structure
\be
     V = V_N \otimes H^2 \ .
\label{space}
\ee
The $N$-dimensional space $V_N$ is multiplied by a countably infinite 
dimensional space $H^2$ associated with the countable infinity of copies 
of D0-branes and strings stretching between them\footnote{Recently, 
winding supermembranes in eleven dimensions were investigated in \cite{WPP}.}.

The description of the Matrix model becomes simpler upon performing
T-duality in directions $X_1,X_2$. The torus $T^2$ is replaced
by its dual torus $\hat{T}^2$, and the 0+1 dimensional quantum mechanics
becomes a 2+1 dimensional U(N) super-Yang-Mills 
theory \cite{GRT,BFSS,Suss}.

Instead of analyzing the fluctuations of a wrapped transverse membrane
in the T-dual 2+1 dimensional Matrix theory, we will try 
to recover a Lagrangian for the massless fluctuations directly from 
the 0+1 dimensional theory on the initial torus $T^2$. 
Our approach will be to construct a membrane on the covering space and  then
to demand periodicity of the spatial coordinates. We consider a constant
density $\sigma_0$ of D0-branes on the covering space. The total number
of D0-branes is thus infinite, consistent with a nonvanishing value
of the membrane charge, $\tr [X_1,X_2] \neq 0$. The construction 
then proceeds much as before: we represent the membrane background
$U_{1,2}$ as covariant derivatives with a U(1) magnetic field, acting
on functions of the membrane world volume,
\bea
     U_1 &\leftrightarrow & 2\pi \zcal (i\pat_1 + a_1)  \\
     U_2 &\leftrightarrow & 2\pi \zcal (i\pat_2 + a_2) \nonumber
\eea 
where ${\bf a} = (a_1,a_2) = -\half f_{12} (x_2,-x_1) $ as before.
However, now the coordinates $x_1,x_2$ have a finite range:
\be
    0 \leq x_r \leq  L_r \ \ , \ r=1,2
\ee
where $L_1,L_2$ are the sizes of the torus $T^2$, and
we demand that all quantities are periodic up to gauge transformations.
With this range for the coordinates $x_1,x_2$, the membrane wraps once
around the torus. The trace is represented as in (\ref{trace1}), leading 
to a finite value of total charge 
\be
   Z_{12} = -\frac{i}{2\pi} \ \sigma_0 \ \int \ d^2x \ [U^1,U^2]
          = \sigma_0 \zcal L_1L_2 = \frac{L_1L_2}{(2\pi)^2} \ .
\label{2charge}
\ee
for the wrapped membrane.

The fluctuation analysis proceeds as before, and leads to the same effective
2+1 dimensional Lagrangian (\ref{U(1)L}) as before. 
The tension of the membrane can be checked
to be given by the relation (\ref{tension}), as it should.

Consider now 2 parallel membranes wrapped on $T^2$. In addition to 
the excitations of open strings within each membrane, there are excitations
of strings which interpolate between the two membranes. 
As discussed in \cite{HT}, these strings 
do not have winding modes because they
are all homotopic to each other. Effectively, these strings behave as
if the spacetime would be non-periodic. However, the homotopy property
and the periodicity of the torus is seen in the possibility of introducing
Wilson lines which correspond to non-trivial gauge holonomy.
We will return to this issue shortly. 

For two parallel membranes, we could write
\bea
     U_1 &\leftrightarrow & 2\pi \zcal \ (i\pat_1 + a_1) \otimes {\bf 1}_2 \\
     U_2 &\leftrightarrow & 2\pi \zcal \ (i\pat_2 + a_2) 
                                                  \otimes {\bf 1}_2 \nonumber
\label{twosimple}
\eea
as before, and then proceed exactly as we did in the previous section.
This leads to the 2+1 dimensional U(2) Yang-Mills Lagrangian (\ref{U(k)L}).
However, it should be possible to make distinction between two singly
wound (around $T^2$) membranes and a {\em single} membrane winding
twice around one of the cycles of $T^2$. This distinction can be made
as follows. We can add constant terms to (41) and define
\bea
     U_1 &\leftrightarrow & 2\pi \zcal \ [(i\pat_1 + a_1) \otimes {\bf 1}_2 
     + \bra A_1 \ket ] \\
     U_2 &\leftrightarrow & 2\pi \zcal \ [(i\pat_2 + a_2) \otimes {\bf 1}_2
     + \bra A_2 \ket ] \ . \nonumber
\eea
The constant terms $ \bra A_r\ket ,\ r=1,2$ are U(2) Lie algebra valued. This
addition does not affect the total 2-brane charge $Z_{12}$, since 
the commutator $[\bra A_1\ket ,\bra A_2\ket ]$ is 
traceless under $\tr_2$. However,
in order to leave the 2-brane charge {\em density} unaffected, we 
must require that the $\bra A_1 \ket $ and $\bra A_2\ket $ commute.

Recall that the fluctuations $A_1,A_2$ around the background are 
defined by
\bea
     X_1 = U_1 +  A_1 &\leftrightarrow & 
             2\pi \zcal \ [(i\pat_1 + a_1) \otimes {\bf 1}_2
           + \bra A_1\ket + A_1 ] \\
     X_2 = U_2 +A_2  &\leftrightarrow & 
             2\pi \zcal \ [(i\pat_2 + a_2) \otimes {\bf 1}_2
           + \bra A_2\ket  + A_2 ]  \ . \nonumber
\eea
{}From these relations it is apparent that the constant terms $\bra A_r\ket $ 
can be interpreted as constant backround values of the gauge fields $A_r$ 
in the U(2) Yang-Mills Lagrangian (\ref{U(k)L}). Thus, they can be
used \cite{HP} to induce gauge 
holonomies $\holo_r$, through the Wilson lines
$$
     \holo_r = P\exp \{ i\oint_r \ A \} 
$$
around the cycles of $T^2$. When the background values are both zero, the
holonomies are trivial,
$$
  \holo_1 = \holo_2 = {\bf 1}_2 \ , 
$$
around both cycles. From the IIA point of view, we then have a bound
state of two D2-branes (and D0-branes), both winding once around the
two cycles of $T^2$. But if we set
\bea
     \bra A_1\ket &=& \frac{\pi}{2L_1} \left( \begin{array}{cc} -1 & 1 \\ 
                1 & -1 \end{array} \right) \\ \nonumber
     \bra A_2\ket &=& 0 \ , 
\eea
we get a non-trivial holonomy
$$
  \holo_1 = \left( \begin{array}{cc} 0 & 1 \\ 1 & 0 
                   \end{array} \right) \ .
$$
In this case,
the result is interpreted as a single D2-brane (with D0-branes) which
winds twice around the $X_1$ cycle of $T^2$.
Both cases yield the same 2-brane charge $Z_{12}$ which is twice the
charge (\ref{2charge}) of a single membrane. Generalizations to other
winding membranes can be obtained in similar manner. 

\bigskip

\subsection{Longitudinal fivebrane backgrounds (p=4)} 

\bigskip

The fivebrane of M theory admits a simple description in Matrix theory
only if it is wrapped around the longitudinal direction. Since the fivebrane
is boost invariant in its world volume directions, going to the infinite 
momentum frame is more involved than for the transverse membrane.
Momentum is instead added by superimposing
a gravitational wave solution. However, in the case of a single fivebrane
such a configurations necessarily includes nonzero membrane charge as well.
To obtain a configuration with vanishing membrane charge one can take 
two fivebrane solutions with mutually opposite membrane charges and then
combine them. In this section we will examine fluctuations around
both types of backgrounds. First, let us 
summarize the backrounds and their properties: 

\begin{description}

\item [(i)] an infinite longitudinal fivebrane with transverse membrane charge 
density; this reduces in D=10 to a type IIA non-marginal bound 
state $4+2\perp 2+0$

\item [(ii)] infinite longitudinal fivebranes 
with transverse leftmoving 
oscillations carrying momentum density in 11th direction; this reduces 
in D=10 to a marginally bound $4 \parallel 0$  configuration

\end{description} 

\noindent
In case (i), in the $\ninfty$ limit the background is represented by
$$
 U_r \repres \left\{ \begin{array}{l}
               2\pi \zcal \ (i \pat_r + a_r )  \ \ \ r=1,2 \\ 
               2\pi z_{34} \ (i \pat_r + a_r )  \ \ \ r=3,4 
             \end{array} \right.
$$
where $a_r$ is a U(1) background field with a field 
strength $f_{rs} = \pat_r a_s - \pat_s a_r$,
with nonzero constant components $f_{12},f_{34}$ describing 
magnetic flux densities in 1-2 and 3-4 planes,
$$
 f_{12} = \frac{1}{2\pi \zcal} \ \ , \ \ f_{34} = \frac{1}{2\pi z_{34}} \ . 
$$
A similar
analysis as in Section 2.1 yields an effective Lagrangian for the 
bosonic fluctuations,
\be
 {\cal L}_B = \sigma_0 \ \int \ d^4x \ \frac{T_0}{2} \{ -\frac{1}{2} 
 (F_{\mu \nu} F^{\muhat \nuhat}) + \pat_{\mu} \phi_i \pat^{\muhat} \phi_i 
    - f_{rs} F^{\rhat \shat} 
    - \half (f_{rs} f^{\rhat \shat}) \}
\label{U(1)5L}
\ee
where the indices are raised with the metric
\be
    \hat{\eta}_{\mu \nu } = {\rm diag} 
                         \ (1,-f^2_{12},-f^2_{12},-f^2_{34},-f^2_{34} ) \ , 
\ee
and where $\sigma_0$ is the D0-brane density. 

\noindent
In case (ii), the background is represented by covariant derivatives with
a selfdual U(2) background\footnote{For earlier studies of such
background configurations in non-abelian gauge 
theories, see {\em e.g.} \cite{LVB}.} field:
\be \begin{array}{ll} a_1 =0 & a_3 =0 \\
                a_2 = F_0 x_1 \sigma^3 & a_4 = F_0 x_3 \sigma^3 \ \ .
    \end{array}
\label{double5brane}
\ee
The $\pm 1$ diagonal elements of $\sigma^3$ represent the superposition of 
two solutions carrying opposite membrane charge. Note that the background
breaks the U(2) symmetry to U(1)$^2$. 
The field strength $f_{pq}=\pat_p a_q - \pat_q a_p -i[a_p,a_q]$
satisfies the selfduality condition 
\be
  f_{pq} = \tilde{f}_{pq}  = \half \epsilon_{pqrs} \ f_{rs} \ .
\ee
The transverse volume of the fivebranes 
becomes infinite in the $\ninfty$ limit and
is given by $A=(2\pi )^4 Nz_{1234}$, where $z_{1234}$ is the longitudinal
fivebrane charge density. The D0-brane charge density $\sigma_0$ is
\be
  \sigma_0 = \frac{N}{A} = \frac{1}{(2\pi )^4 z_{1234}} \ .
\ee
It is related to the background field by
\be
 2 \ \sigma_0 = \frac{1}{8\pi^2} \tr_2 \ (f\wedge f) \ ,
\label{wedge1}
\ee
which gives
\be
            F_0 = \frac{1}{2\pi \sqrt{z_{1234}}} \ .
\ee
Note that  $\sigma_0$ is the density of D0-branes per fivebrane; the total
density of D0-branes is  $2 \sigma_0$.

We then represent the background (ii) by the covariant derivatives
$$
 U_r \repres 2\pi \sqrt{z_{1234}} 
            \ (i\pat_r {\bf 1}_2 + a_r) \ \ \ r=1,2,3,4 \ .
$$
We can check that the
configuration carries a longitudinal 5-brane charge density
\be
   -\frac{1}{8\pi^2} \epsilon_{pqrs} 
                                      \tr [U_pU_qU_rU_s] =  2\cdot z_{1234} \ ,
\label{5charge}                 
\ee
where the prefactor 2 represents the presence of two fivebranes. 
The fluctuations $A_0,A_r,\phi_i$ about the background are represented
by U(2) Lie algebra valued fields in 4+1 dimensions:
\bea
    A_0 &\leftrightarrow & A^0_0(x)\  \unittwo + A^a_0(x)\  T^a  \\  
   A_r &\leftrightarrow & 2\pi \sqrt{z_{1234}} 
          \ (A^0_r(x)\ \unittwo + A^a_r(x) \ T^a) \nonumber \\
 \phi_i &\leftrightarrow & \phi^0_i(x)\ \unittwo + \phi^a_i(x)\ T^a 
     \ . \nonumber
\eea 
where $T^a$ are SU(2) generators. The effective action 
for the fluctuations is found to be
\be
 {\cal L}_B = \sigma_0  \ \int \ d^4x \ \frac{T_0}{2} \ \tr_2 \ \{ 
 - \half (F_{\mu \nu} F^{\muhat \nuhat}) 
    + D_{\mu} \phi_i D^{\muhat} \phi_i + \half [\phi_i,\phi_j ]^2
 - f_{rs} F^{\rhat \shat} - \half (f_{rs} f^{\rhat \shat}) \} \  \ ,
\label{U(2)5L}
\ee
where the indices are raised with the metric
\be
 \hat{\eta}_{\mu \nu } = {\rm diag } \ (1,-F^2_0,-F^2_0,-F^2_0,-F^2_0) \ .
\ee
The energy density of the configuration is the sum of fivebrane plus
gravitational wave contributions $H= T+p_{11}$. The infinite momentum
frame Hamiltonian is found by subtracting $p_{11}$: $H_{\rm IMF} = T$.
T is found from the constant term in (\ref{U(2)5L}), 
\be
  T = 2 T_4 = 2 \frac{1}{(2\pi )^2} T_0 \ .
\ee
This is as expected, since two longitudinal fivebranes are interpreted
as two D4-branes in the IIA theory.

\bigskip

\section{Scattering of branes}

Recently, there have been several studies of brane-brane scattering in
the context of Matrix theory \cite{AB,LM,L2,BC,L3,P,BB,CT1,CT2,PP}. These 
calculations have demonstrated
agreement\footnote{Subtler issues and possible
discrepancies have also been discussed, see \cite{BL,GGR}.} 
between 11 dimensional supergravity and Matrix theory 
at one loop and even two loops \cite{BB}.
For Matrix theory compactified on torii, such
calculations amount to loop diagrams in Super-Yang-Mills theory
on the dual torus with magnetic backround fields encoding the brane
configuration and Higgs field VEVs specifying the kinematics of the 
scattering problem. From the discussion in Section 2, it should be obvious
that a similar correspondence applies also for scattering 
in uncompactified Matrix theory involving infinite branes. 

Note that all the effective 
actions (\ref{U(1)L},\ref{U(k)L},\ref{U(1)5L},\ref{U(2)5L}) can be 
rewritten in a form
where the magnetic background fields are combined with the fluctuation
fields. Introducing a U(k) gauge field $\Acal_{\mu}$,
\bea
  \Acal_0 &=& A_0 \\
  \Acal_r &=& a_r + A_r \nonumber
\label{fullfield}
\eea
and denoting its field strength by $\Fcal_{\mu \nu}$, the 
actions (\ref{U(1)L},\ref{U(k)L},\ref{U(1)5L},\ref{U(2)5L}) can
be written in the form
\be
  {\cal L}_B = T_0 \sigma_0 \ \int \ d^px \ \tr_k
\ \{ -\frac{1}{4} (\Fcal_{\mu \nu } \Fcal^{\muhat \nuhat}) 
   + \half ({\cal D}_{\mu} \phi_i {\cal D}^{\muhat} \phi_i )
  + \frac{1}{4} [\phi_i ,\phi_j ]^2 \}
\label{YMaction}
\ee
where the covariant derivative ${\cal D}_{\mu}$ is taken with respect
to the field (57) which includes the background. 
Now, after including fermions, gauge fixing and ghost terms,
we can compute scattering amplitudes from loop diagrams in the 
appropriate background fields which encode the 
information about the scattering objects and kinematical setup.

\subsection{D0 - D6+D0 Scattering}

We will perform a scattering calculation involving configurations 
of D6-branes and D0-branes. Among various other configurations,
G. Lifschytz studied the scattering of 6+4+2+0 bound states from
D0-branes \cite{L2}. (Further discussion can be found in \cite{BL}.)  
Recently, W. Taylor \cite{WT} showed 
how to compose configurations of D6 branes
and D0-branes which do not carry D4 or D2 brane charges. 
Such configurations\footnote{These
configurations are classically stable up to quadratic order. 
They are thought to be related
to the supergravity black hole solutions of \cite{Shein} carrying
0-brane and 6-brane charges.} carry 
a background U(4) gauge field 
\be \begin{array}{lll} a_1 =0 & a_3 =0 & a_5 =0 \\
                a_2 = F_0 x_1 \mu_1 & a_4 = F_0 x_3 \mu_2
                 & a_6 = F_0 x_5 \mu_3
    \end{array}
\ee
with traceless U(4) matrices
\bea
   \mu_1 &=& {\rm diag} (1,1,-1,-1) \\ 
   \mu_2 &=& {\rm diag} (1,-1,-1,1) \nonumber \\
   \mu_3 &=& {\rm diag} (1,-1,1,-1)  \ . \nonumber
\label{mus}
\eea
Analogously to (\ref{wedge1}), the above configuration gives a 
D0-brane charge density
$$
   4\ \sigma_0 = \frac{1}{48\pi^3} 
    \tr_4 \ ( f \wedge f \wedge f) = 4\ \left( \frac{F_0}{2\pi } \right)^3 \ .
$$
As before, one can examine the fluctuations about this configuration
and obtain an effective U(4) action of the form (\ref{YMaction}) with $p=6$ and
a metric
$$
 \hat{\eta}_{\mu \nu } = {\rm diag} (1,-F^2_0,-F^2_0,
                                            -F^2_0,-F^2_0,-F^2_0,-F^2_0) \ .
$$
We have examined
the scattering of D0 particles from Taylor's D6+D0 brane configurations and
calculated the potential between  these objects. 
The D6+D0 configuration can be understood in the same way as the
fivebrane with vanishing membrane charge of Section 2.3. It represents
a superposition of four 6+4+2+0 solutions which are chosen to give 
vanishing D4-brane and D2-brane charges when combined. 
The scattering calculation that follows reveals this structure, in
as much as the potential is found to be four times that between
a D0 particle and a 6+4+2+0 state.
In the Matrix theory 1-loop calculation
(which is the same as the 1-loop calculation in the effective 
theory (\ref{YMaction}),
as we argued above), we encounter the following determinants
(the boson, ghost and fermion contributions in the background 
gauge) 
\be \begin{array}{l}
 \det^{-3} \{ (-\pat^2_{\tau} + H +2c) \ {\bf 1}_4  \} 
 \ \det^{-3} \{ (-\pat^2_{\tau} + H -2c) \ {\bf 1}_4  \} \\ 
 \det^{-2} \{  (-\pat^2_{\tau} + H ) \ {\bf 1}_4 \} \\
 \det \{  (-\pat^2_{\tau} + H +3c) \ {\bf 1}_4 \} 
 \ \det^3 \{  (-\pat^2_{\tau} + H +c) \ {\bf 1}_4  \} \\
 \det^3 \{  (-\pat^2_{\tau} + H -c) \ {\bf 1}_4  \}
 \ \det \{ (-\pat^2_{\tau} + H -3c) \ {\bf 1}_4  \} \ \ ,
\end{array}
\ee
where
$$
 H = r^2 + c (2(n_1+n_2+n_3)+3)  \ .
$$
Here $r$ is the distance between the D0-brane and 
the 6+0 configuration, $c=1/F_0$,
and $n_1,n_2,n_3=0,1,\ldots$ . These determinants
are the same as those in the calculation by Lifschytz
of the D0 - 6+4+2+0 potential, except for additional
$4\times 4$ unit matrix factors. The unit matrices signal that the D0-brane
sees the four 6+4+2+0 sublayers in the 6+0 configuration, thus the
potential between the 6+0 configuration and the D0-brane will be four times
that obtained by Lifschytz \cite{L2}:
\be
     V(r) = 4 \cdot \frac{3}{16} \frac{1}{F_0r} \ .
\label{potential}
\ee

The corresponding supergravity result is easily obtained
by considering the 6+0 configuration as a probe moving 
in the D0-brane background. The calculation  is performed in the
same manner as by Chepelev and Tseytlin \cite{CT1}, who considered
various configurations involving 4-branes and 5-branes. Their paper
also collects a number of useful formulae which we will draw from in
the following. The effective action for the 6+0 configuration is 
given by\footnote{For the effective action we will use Tseytlin's
truncated version of the non-abelian Born-Infeld action \cite{NABI}, 
valid in cases,
such as the present one, where the worldvolume field strengths commute.}
\be
 S_6 = \int \ d^7 \xi \ \tr \{ -T_6 e^{-\phi } \ \sqrt{\det [g_{\mu \nu}
\frac{\pat X^{\mu}}{\pat \xi^i} \frac{\pat X^{\nu}}{\pat \xi^j}
 + F_{ij}]} + \mu_6 (C_0 F_{12}F_{34}F_{56} + \cdots ) \} \ ,
\ee
where $\cdots$ indicates various Chern-Simons terms which are irrelevant
for our present purposes. In our units, we have $\mu_6 =T_6$. The field
strength in the worldvolume is 
given by $F_{12}=F_0\mu_1,F_{34}=F_0\mu_2,F_{56}=F_0\mu_3$, with 
$\mu_{1,2,3}$ defined in (60). The spacetime fields to be inserted
into $S_6$ are those induced by a D0-brane source ``smeared'' in directions
$x^1,\ldots ,x^6$:
\bea
ds^2 &=& H^{-1/2}_0 dt^2 - H^{1/2}_0 
      (dx^2_1 + \cdots +dx^2_9) \\ \nonumber
e^{-\phi} &=& H^{-3/4}_0 \ \ ; \ \ C_0~=~H^{-1}_0 -1 \\ \nonumber
H_0 &=& 1 + \frac{Q^{(6)}_0}{r} \ \ ; \ \ 
                     r^2~=~x^2_7 + x^2_8 + x^2_9 \\ \nonumber
Q^{(6)}_0 &=& \frac{g}{2} (2\pi )^{5/2} \ .
\eea
Substituting into $S_6$ yields 
$$
  S_6 = -4T_6 \int \ d^7x \ \{ H^{-1}_0 (H_0 + F_0^2)^{3/2} 
                     - (H^{-1}_0 -1)F_0^3 \} \ .
$$
To compare with the Matrix theory result we should 
take $r,F_0 \rightarrow \infty $, corresponding to the large distance interaction
between objects in the infinite momentum frame of M theory. Then
$$
 S_6 \approx -4T_6 \int \ d^7x \ \{ \frac{3}{8} \frac{H_0}{F_0} 
   + F_0^3 + \frac{3}{2} F_0 \} \ .
$$ 
Now we can read off the potential, 
$$
  V(r) \approx 4 \cdot \frac{3}{8} \frac{T_6 Q^{(6)}}{F_0r} 
   = 4\cdot \frac{3}{16} \frac{1}{F_0r} \ ,
$$
in agreement with the Matrix theory result (\ref{potential}).
 
\section{Conclusion}

We have seen that by taking the $\ninfty$ limit of Matrix theory in a 
particular way, it is possible to recover much of the known physics of 
D-branes in IIA theory.  In particular, by representing the coordinate
matrices as covariant derivatives we arrived at the correct Yang-Mills 
actions governing the low energy dynamics of D-branes.  As discussed, these
actions are properly interpreted as  Born-Infeld actions expanded around 
large magnetic field backgrounds.  This interpretation was verified 
explicitly, including checking numerical factors, in the case of a single
D2-brane.  We also used the effective actions to compute the potential
between a D0-brane and a configuration of D6-branes and D0-branes, and saw
that the result was in agreement with supergravity.    
  
Finally, although in this paper attention has been restricted to 
the $\ninfty$ limit, it would be interesting to investigate D-brane actions at 
finite N. Despite the absence of infinite BPS branes for such N, it is
 possible to compare various processes with the predictions of 
supergravity compactified along a null direction \cite{Nfin,BBPT}.  To understand
such processes better, it would be useful see
 to how much of our techniques may be carried over to finite N.

\bigskip

{\Large {\bf Acknowledgements}}

\bigskip

We would especially like to thank Soo-Jong Rey for discussions which
motivated this investigation. We also thank A.~Tseytlin
for remarks on Matrix theory scattering calculations. 
E.~K-V. would
like to thank the Helsinki Institute of Physics for hospitality
during the writing of this work.

\end{document}